# Probing the Griffiths like phase, unconventional dual glassy states, giant exchange bias effects and its correlation with its electronic structure in $Pr_{2-x}Sr_xCoMnO_6$

Arkadeb Pal[1], Prajyoti Singh[1], Vinod K Gangwar[1], Amish G Joshi[2], G. D. Dwivedi[3], Prince K Gupta[1], Md. Alam[1], Khyati Anand[1], Anup K Ghosh[4], and Sandip Chatterjee[1#]

[1]Department of physics, Indian Institute of Technology (BHU) Varanasi

[2]CSIR-National Physical Laboratory, Dr. K. S. Krishnan Road, New Delhi 110012, India

[3]Department of physics, National Sun Yat-sen University, Taiwan

[4]Department of physics, Banaras Hindu University

Varanasi 221005

Corresponding author's email address: schatterji.app@iitbhu.ac.in

## Abstract

Electronic structure, electrical transport, dc and ac magnetization properties of the hole substituted ($Sr^{2+}$) partially B-site disordered double perovskite $Pr_{2-x}Sr_xCoMnO_6$ system have been investigated. Electronic structure was probed by employing X-ray photoemission spectroscopy (XPS) measurements. The study suggested the presence of mixed valence states of the B-site ions ($Co^{2+}/Co^{3+}$ and $Mn^{3+}/Mn^{4+}$) with significant enhancement of the average oxidation states due to hole doping. The mere absence of electronic states near the Fermi level in the valence band (VB) spectra for both of the pure (x=0.0) and Sr doped (x=0.5) systems indicated the insulating nature of the samples. Sr substitution is observed to increase the spectral weight near the Fermi level suggesting for an enhanced conductivity of the hole doped system. The temperature variation of electrical resistivity measurements revealed the insulating nature for both the systems, thus supporting the VB spectra results. The resistivity curves were observed to follow the variable range hopping (VRH) mechanism in the entire temperature range while the analysis showed a significance enhancement in the carrier concentration due to the hole doping. The dc magnetization data divulged a Griffiths like phase above the long range ordering temperature. A typical re-entrant spin glass like phase driven by the inherent anti-site disorder (ASD) has been maidenly recognized by ac susceptibility study for both the pure and doped systems. Most interestingly, the emergence of a new cluster glass like phase (immediately below the magnetic ordering temperature and above the spin-glass transition temperature) solely driven by the Sr substitution has been unravelled by ac magnetization dynamics study. Observation of these dual glassy states in a single system is scarce and hence placed the present system amongst the rare materials. The isothermal magnetization measurements further probed the exhibition of the giant exchange bias effect emanated from the existence of multiple magnetic phases.

## INTRODUCTION

The technology of the next generation spintronic devices are progressively predicated upon discovering new multifunctional materials that respond to multiple external stimuli. Such multifunctional properties of strongly correlated oxide systems typically emanate from the existence of coupled microscopic degrees of freedom viz., spin, phonon, and polarization etc,. Hence, the search for such materials has got invigorated interest in recent past so that their novel properties can be harnessed to fabricate innovative practical applications such as memory devices, spin-filters, spin switching, and capacitive devices etc [1-5]. Among such emerging multifunctional materials, a great deal of research attention have been bestowed on the class of double perovskites (DP) $A_2BB'O_6$ (A= Rare earth ions or alkaline ions; B/B$'$= transition metal ions) owing to their wide range of exotic properties including near room temperature ferromagnetic ordering, colossal magneto-resistance, giant magneto-dielectric coupling, magnetocaloric effects, Griffiths phase, giant exchange bias, spin glass, spin-phonon coupling, metamagnetic transition etc [6-12]. Apart from their far reaching technological prospects, the DPs offered a fascinating playground for the basic solid state research towards the understanding of the competing interplay of the magnetic, electric and phononic order parameters [12-14]. In its ordered structure, the two sub-lattices consisting of $BO_6$ and $B'O_6$ octahedrons form a rock-salt type ordering [15-16]. The novel and unconventional ground states of such DPs are mainly established by the B-site cationic ordering and their nominal valence states. Most of the Co/Ni/Mn based ordered DPs exhibit ferromagnetic (FM) insulating ground state which is best understood by the $180^0$ positive super-exchange interactions among $B^{2+}/B'^{4+}$ ions via adjacent oxygen ions following Goodenough-Kanamori rules [16-17]. As a matter of fact, though a perfectly ordered structure offers a desirable platform for studying its properties theoretically but is extremely difficult to synthesize. The disorders in the form of anti-site disorder (ASD) i.e. site exchange between B/B$'$ ions inevitably creeps into the structure while sample preparation. Additionally, introduction of $B^{3+}/B'^{3+}$ ions as a disorder leads to competing anti-ferromagnetic (AFM) interactions in the system [18-19]. It has been observed that such anti-site disorder has profound effects on the physical properties of DPs, especially the magnetic properties and can lead to fascinating phenomena viz., Griffiths phase, exchange bias, low temperature spin-glass states, room temperature giant magneto-dielectricity etc [8, 10, 11]. However, despite of rigorous theoretical and experimental investigations via magnetic, neutron, NMR, X-ray absorption/emission spectroscopy etc studies on such DPs, a complete understanding of the electronic structure and the impacts of the B-site cationic ordering on its properties is far from being well-understood and still remained as an open challenge [19].

The spin-glass (SG) state in a system is a manifestation of the spin frustration due to the existence of competing magnetic interactions leading to the macroscopically degenerate multi-valley ground states [20-22]. In a SG state, the competing spins conflict with each other and cannot attain a long range magnetic ordering, rather freeze in a random non-collinear fashion thus leading to a nonergodic state characterized by aging effect, thermo-magnetic irreversibility, slow spin dynamics etc. Unlike the typical spin glass states, recently glassy spin dynamics have been observed in some systems where the entity responsible for such frustrated non-equilibrium state is not really the atomic spins; instead it is the spin-cluster. Such states are known as the cluster glass (CG) state [23-25]. As a matter of fact, in SG states, the competing FM/AFM interactions tend to have equivalent strengths on lowering the temperature while the CG state evolves if one of the competing interactions (AFM or FM) is weaker relative to the other. However, the existence of the static-disorders in the system is often seen to be responsible for anticipating the short range magnetic ordering such as SG or CG states at lower temperatures in addition to the existing higher temperature long range magnetic ordering, known as re-entrant spin glass (RSG) or cluster glass (RCG) states [26-28]. This particular dynamic magnetic property recently has attracted immense research interest world-wide and remained at the forefront of research. This special phase is realized by the co-existence of the long range order parameter even in the glassy state in the lower temperature. Eventually, as compared to the large number of systems exhibiting SG states, fewer systems are reported to show RSG or RCG states. The different theoretical and experimental investigations showed different possible mechanisms are associated to the observed glassiness viz., site disorder, geometrical spin frustration, spin-orbit interaction, octahedral tilting etc [8, 28-31]. However, the exact origin is still an open challenge and requires more efforts for a complete understanding. Apart from its intriguing rich physics, it helped envisaging many real world practical applications viz., protein folding, memory applications, neutral networks etc [22,32,33].

On the other hand, when a system possesses multiple magnetic phases, an exchange bias can be induced exhibiting a shift along the field axis of the M(H) (hysteresis loop) curves [34-37]. This happens due to the formation of the FM unidirectional exchange anisotropy at the interface of the two different magnetic phases. Since its discovery, it has received tremendous research attention due its scopes for the real life applicability in spin-valves, read heads and ultra-high density magnetic memory devices etc [37,38].

Eventually, previous reports available on the partially disordered DP $Pr_2CoMnO_6$ system revealed its dc magnetic properties and exhibition of strong spin-phonon coupling [39]. However, an extensive study of this system by ac susceptibility measurements deciphering its low temperature spin dynamics is hitherto unreported. Again, to date, no comprehensive reports are available describing its electronic structure. Moreover, hole doping in the manganite perovskites has received tremendous research attention since the discovery of the colossal magnetoresistance in such materials [40]. The underlying physics of such systems involves the phase separation due to the intrinsic competition of spins, lattice, orbital or charge orders leading to the inhomogeneous magnetic and electronic ground states [41]. Understanding the above facts, we got motivated to carry out an extensive study on hole ($Sr^{2+}$) doped DP $Pr_{1.5}Sr_{0.5}CoMnO_6$ system via detailed dc and ac magnetization, transport, X-ray photoemission spectroscopy (XPS) studies. The possible impacts of the inherent B-site disorder and as well as the heterovalent ion ($Sr^{2+}$) substitution on its electronic structure and physical properties have been thoroughly investigated.

## EXPERIMENTAL DETAILS

**Synthesis:**

The synthesis of the polycrystalline $Pr_{2-x}Sr_xCoMnO_6$ (x=0.0, 0.3 and 0.5) samples were done following the standard solid state reaction route. The highly pure (>99.99%) oxide precursor powders of $Pr_6O_{11}$, $SrCoO_3$, CoO and $Mn_2O_3$ were taken in proper stoichiometric ratio followed by intimate grinding for an hour. Initial heat treatment of this powder was done at $950^0$ C for 24 hours in air. After regrinding, the powder was subjected to several heating cycles at $1150^0$C with reheating and intermittent grindings for several days. Finally, the powder thus obtained was pelletized and sintered at $1250^0$ C for 36 hours followed by a slow cooling to room temperature.

**Characterizations:**

X-ray diffraction measurements at 300 K were carried out using Rigaku Miniflex II X-ray diffractometer. The Rietveld refinements of the X-ray diffractograms (XRD) of the samples confirmed that all the three samples crystallized in single phase monoclinic structure with $P2_1/n$ symmetry (not shown). The superconducting quantum interference device (SQUID) based magnetic property measurement system (Quantum Design-MPMS) was employed for magnetization measurements. The XPS data was recorded by an Omicron multi-probe surface science system which is well-equipped with a hemispherical electron energy analyzer (EA 125) along with a monochromatic X-ray source Al-$K_\alpha$ line with photon energy 1486.70 eV. The average base pressure was maintained at a value of ~$2.8\times10^{-11}$ Torr. The total energy resolution as calculated from the width of Fermi edge was about 0.25eV.

## Results and discussions

**X-ray photoemission spectroscopy study:**

To understand the magnetic and transport properties of the present systems, it is necessary to have a prior knowledge of its electronic structures. In order to do so, we have performed X-ray photoemission spectroscopy (XPS) study of systems x=0.0 (PCM) and 0.5 (PCMS25). In XPS study of a system containing open-shell ions, the coupling between the open-shell and a core electron vacancy produces the multiplet structure. Apart from the features of the main photoelectron lines of XPS, the pertaining satellite peaks, relative intensities of the core level peaks and chemical shifts are helpful to precisely determine the ligand co-ordinations and nominal charge states etc [42]. All peak positions and doublet separations of the recorded data have been assigned from the National Institute of Standards and Technology (NIST) XPS database. Survey scan confirms the presence of Pr, Co, Mn and O in both the samples, while Sr is present only in PCMS25 sample (Fig. 1 (a)). However, no trace of any other impurity elements except C was detected from the spectrum. The presence of C*1s* peak is related to the adventitious molecules which got absorbed at the surface from the air. Core-level XPS of Pr*3d* region of both pure (PCM) (upper panel) and 25 %Sr doped (PCMS25) (lower panel) show spin-orbit split peaks Pr$3d_{3/2}$ and Pr$3d_{5/2}$ around 953 eV and 933 eV respectively, with doublet separation of 20.4 eV (Fig. 1(b)). Two additional exchange splitting features positioned around ~4.8 eV below the main Pr3d peaks can also be observed which are originated from the coupling of Pr*4f* and Pr*3d* hole states [42,43]. Moreover, a shoulder above the Pr$3d_{3/2}$ peak marked as "t" is also observed which is seemingly associated to the multiplet effect. All these features suggest for the trivalent oxidation states of the Pr ions [42,43].

Core-level XPS of Mn*2p* region of PCM and PCMS25 samples are shown in Fig. 1(e). For both the samples, the spin-orbit splitting (ΔE) of the peaks Mn$2p_{1/2}$ and Mn$2p_{3/2}$ are found to be around ~11.7 eV. In fact, ΔE for Mn*2p* XPS spectra in $MnO_2$ is around ~11.8 eV [44]. The reduced value of ΔE~11.7 eV for the present samples suggests that valence state of Mn in both systems is slightly smaller than +4, which is possibly due to the presence of fractional $Mn^{3+}$ ions. The analysis by deconvolution of the Mn2p XPS spectra yielded that $Sr^{2+}$ ion substitution slightly increased the concentration of $Mn^{4+}$ as compared to the pure sample. This little increase in $Mn^{4+}$ concentration was expected as after heterovalent $Sr^{2+}$ ion (hole) substitution, the system would essentially try to maintain the charge neutrality by causing an effective increase in the Co/Mn ions oxidation states.

The XPS study of the compounds containing Co is particularly important since its spectra contain shake-up satellite peaks (due to poorly screened states) along with the main photoelectron lines (well screened states) which are highly sensitive to oxidation states and ligand co-ordination etc [42,45,46]. In Fig. 1(f), the core-level Co*2p* XPS spectra of PCM (top) and PCMS25 (bottom) comprising of spin-orbit split Co$2p_{1/2}$ and Co$2p_{3/2}$ peaks are shown. Eventually, the Co*2p* doublet separation (ΔE) for PCM is found to be ~15.7 eV while for PCMS25 it comes down to 15.4 eV. In fact, Co*2p* doublet separation in CoO (divalent Co) and $Co_3O_4$ (mixed +2 and +3 states) are reported to be ~15.9 eV and 15.3 eV respectively [42,46]. This suggests that for both the PCM and PCMS25 samples, the Co cations exist in mixed valence +2/+3 states. However, the observed decrease in doublet separation results from the increased oxidation states of Co due to Sr doping. Moreover, it is pertinent to note that two intense charge transfer satellite features are observed above the main Co*2p* peaks in PCM which are relatively weak in PCMS25. Such satellite features nearly ~7 eV above Co*2p* main peaks are typically observed in CoO and thus used as a hallmark feature for $Co^{2+}$ ions [42,45,46]. In contrast, these are nearly absent or very weak in Co*2p* XPS spectra of $Co_2O_3$ or $LiCoO_2$ where Co ions exist in +3 states. Whereas for $Co_3O_4$, the Co*2p* XPS spectra exhibits weaker satellites due to presence of mixed $Co^{2+}$/$Co^{3+}$ valance states [45,46]. Hence, again the reduced satellite intensity for Sr doped sample clearly suggests for increased $Co^{3+}$ ions relative to that in pure sample [42]. The deconvolution analysis of the respective Co*2p* XPS spectra of PCM and PCMS25 also yielded significant increase in the $Co^{3+}$ concentration due to Sr doping.

O*1s* core-level XPS of both systems are shown in Fig. 1(c). Peak around ~529 eV is the representative peak of $O^{2-}$ ions of lattice oxygen [42] and while the peak around ~531 eV~532 eV is attributed fewer electron rich oxygen species (viz., $O_2^{2-}$ , $O_2^{-}$ or $O^{-}$) due to the adsorption of oxygen at the surface [42]. Fig. 1(d) depicts Sr*3d* XPS spectrum for PCMS25 system. The peak positions and the doublet separation of ~1.8 eV between spin-orbit split peaks Sr$3d_{3/2}$ (~133.2 eV) and Sr$3d_{5/2}$ (~131.4 eV) indicate +2 oxidation states for Sr [47].

Fig. 2(g) shows the valence band (VB) spectra recorded at 300 K for both the PCM and PCMS25 samples. Apparently, below the Fermi level ($E_F$), both the spectra are showing similar patterns consisting of two main intense features denoted as A and B which are positioned at ~2 eV and 5.7 eV respectively. The VB spectra for both the samples showed weak spectral weight across the $E_F$, thus indicating insulating nature of the samples. However, for the PCMS25 sample, weak but finite states can still be visible near $E_F$ which predicts relatively enhanced conductivity in it. Eventually, both the VB spectra are mainly composed of hybridized states of Pr*4f*, Co*3d*, Mn*3d* and O*2p* orbitals [42,48]. As a matter of fact, the occupied spectral feature immediately below $E_F$ (0-1 eV) can mainly be ascribed to the hybridization of the extended Co*3d* ($e_g$) and Mn*3d* ($e_g$) states [42,48-51]. The contribution to the most intense feature A is predominantly coming from the Pr*4f*, Co*3d* ($t_{2g}$), Mn*3d* ($t_{2g}$) states hybridized with O*2p* states while the significant proportion being from Pr*4f* states [42]. On the other hand, the next feature B can be attributed mainly to the hybridization of the extended Co*3d* ($t_{2g}$) and Mn*3d* ($t_{2g}$) states with ligand O*2p* states (other minor contribution being from O*2p*-Co/Mn 4*sp* and O*2p*-Pr *5sd* oxygen bonding states.). However, the decreased relative intensity of A for PCMS25 sample is seemingly associated to the decreased Pr*4f* states (due to Sr substitution).

**Electrical resistivity study:**

Fig. 2 represents the electrical resistivity (ρ) variation with temperature for PCM and PCMS25. For both the curves, the resistivity shows monotonous increase with decreasing temperature, thus exhibiting an insulating nature of the samples. However, as evident from the figure, Sr substitution caused a dramatic drop in the resistivity. This can be presumably attributed to the enhanced carrier concentration due to the effective increase in the mean oxidation states of the B-site ions (i.e. Co/Mn) by Sr doping. As a matter of fact, the resistivity behaviour of many double perovskites systems is often observed to follow the 3D Mott variable range hopping (VRH) mechanism (of the form:$\rho = \rho_0 \exp\left(\frac{T_0}{T}\right)^{1/4}$, where $\rho_0$ represents the pre-factor and $T_0$ is related to the characteristic hopping temperature associated to the electron hopping probability) in which the most frequent hopping does not occur to the nearest neighbour unlike simple thermally activated process [52]. Eventually, the resistivity curves for both the PCM and PCMS25 systems were satisfactorily fitted with the VRH law in the entire temperature range (Fig. 2). The best fit yielded values of the constant $T_0$ to be ~$1.59\times10^8$ K and $6.43\times10^6$ K for PCM and PCMS25 respectively. $T_0$ is related to the density of the localized states $N(E_F)$ as $T_0 = \frac{24}{\pi K_B N(E_F)\xi^3}$ ; here ξ is the decay length of the localized wave function [52]. The estimated the values of $N(E_F)$ (taking ξ~0.38 which is the mean distance between the neighbouring Co/Mn atoms) are found to be~$1.12\times10^{19}$/eV-cm$^3$ and ~$2.72\times10^{20}$/eV-cm$^3$ for PCM and PCMS25 respectively. The obtained values of $N(E_F)$ are consistent with other similar oxide semiconductors [52].

**Magnetization study:**

Fig. 3 (a) illustrates the temperature (T) variation of dc magnetization (M) of PCMS25 following the conventional field cooled (FC) and zero-field cooled (ZFC) protocols with applied magnetic field of 100 Oe. The magnetization exhibits a sharp jump around $T_C$ ~153 K which suggests for an onset of magnetic ordering below this temperature. The temperature $T_C$ ~153 K has been precisely identified from the inflection point of "dM/dT Vs T" curve (inset top of Fig. 3a). Moreover, the observation of the large thermo-magnetic irreversibility between the ZFC and FC arms clearly indicates towards the existence spin frustrations owing to the different competing spin interactions. Moreover, another broad anomaly near ~35 K is also observed in the "dM/dT Vs T" curve, thus suggesting for the existence of a secondary magnetic phase at lower temperatures. Furthermore, to estimate the effective paramagnetic moment ($\mu_{eff}$) and the Curie-Weiss temperature ($\theta_{CW}$), standard Curie-Weiss (CW) model: $\chi^{-1} = \frac{H}{M} = \frac{3K_B}{\mu_{eff}^2}(T - \theta_{CW})$, has been employed to fit the "temperature (T) variation of inverse susceptibility ($\chi^{-1}$) curve" in the elevated temperature range (inset bottom of Fig. 3a). The fit yielded $\mu_{eff}$~7.78 $\mu_B$ and $\theta_{CW}$ ~ +113 K. The positive value of $\theta_{CW}$ indicates towards the predominant ferromagnetic interactions in the system. Moreover, the large difference between the ordering temperature ($T_C$) and $\theta_{CW}$ suggests for the spin frustration present in the system. The value of the frustration parameter f=$\theta_C/T_C$~0.73 implies the magnetic frustration arising from disorder unlike the geometrically frustrated systems. However, the theoretical spin-only moment of the system has been estimated to be ~7.54 $\mu_B$ (taking high spin states of the individual ions). The larger value of experimental $\mu_{eff}$ than the theoretically calculated one is seemingly associated to the short range FM interactions well above $T_C$ [53,54]. Eventually, the observed down turn behaviour of $\chi^{-1}$ above $T_C$ along with the larger value of the experimental $\mu_{eff}$ (than theoretically expected) is typically observed in a system exhibiting Griffiths phase (GP) [10,54-59]. We have further studied the temperature variation of $\chi^{-1}$ at different applied fields to investigate the GP (Fig. 3b). It is evident from the figure that the $\chi^{-1}$ curves at lower fields are showing clear down turn behaviour at temperatures >$T_C$. This behaviour is typically considered as a hallmark feature of GP as in this regime, the magnetization fails to behave like an analytical function of magnetic fields thus deviating from CW law [58,59]. This basically emanates from nucleation of small but finite sized correlated clusters having short range magnetic ordering embedded in the global paramagnetic matrix above $T_C$. Therefore, the GP is a special and peculiar magnetic phase where the system neither behaves like a paramagnet nor shows long range ordering. Moreover, it can be also observed from the figure that the down-turn deviation gets softened with increasing magnetic fields and yields CW like behaviour which is also an important characteristic of GP. The higher fields eventually raise the background PM moments and consequently the FM cluster contribution gets masked by it, thus obliterating the down-turn feature. The Griffiths phase temperature ($T_G$) (referring to the highest temperature at which short range spin correlations exist and above which the system enters in a perfectly PM region) has been identified to be ~163 K as the down-turn deviation from CW law commences below this temperature (Inset top of Fig. 3b) [54-59]. In the GP regime i.e. $T_C<T<T_G$ (shaded region of the inset top of Fig. 3b), the susceptibly becomes non-analytic function of the magnetic field due to low density FM cluster formation and at lower fields it follows the power law with a characteristic non-universal exponent λ describing Griffiths singularity [57-59]:

$$\chi^{-1}(T) \propto (T-T_c^R)^{1-\lambda}, \ (0<\lambda<1)$$

Here, λ is a measure of deviation from CW behaviour and $T_c^R$ is the reduced long range ordering temperature due to random non-magnetic dilution [55-57]. Thus to confirm the GP, we have used the above formula for fitting the $\log_{10}$-$\log_{10}$ plot of $\chi^{-1}$ Vs $(T-T_c^R)$ for our data at 100 Oe (inset bottom of Fig. 3b). Though a choice of $T_c^R=\Theta_{CW}$ would lead to λ~0 at PM region, however for the present case, due to the co-existence of FM/AFM interactions $\Theta_{CW}$ lies much below of FM ordering temperature. In such cases, to overcome this confusion, a reasonable choice is $T_c^R \sim T_{c2}$ [59]. The obtained λ value of ~0.89 from the linear fit in the GP region suggests the presence of GP in the sample. However to elucidate the observed GP, we may consider the role of the quenched disorder in the system since it is a key ingredient to bring forth GP by hindering the long range ordering and favouring the nucleation of the short range ordered FM clusters [60]. Thus, it is plausible to state that the origin of GP mainly lies in the different sources of quenched disorder in the present system PCMS25 such as the inherent anti-site disorder at B-sites (ASD) and the A-site disorder due to random substitution of the $Sr^{2+}$ ions (different than $Pr^{3+}$ both in size and oxidation states) giving rise to the different competing interactions [61-62]. Moreover, presence of Jahn-Teller (J-T) active ions causing the static quenched disorder is also reported to raise GP which can also be a possible factor contributing towards the observed GP in the present system (as $Co^{2+}$ and $Mn^{3+}$ ions are J-T active) [63]. Contradistinctively, Salaman et al have explained the onset of GP in perovskite $La_{1-x}Ca_xMnO_3$ ($x\rightarrow 0.3$) due to the changes in the Mn-O-Mn bond angles by smaller ion substitution [57]. Hence, the substitution of significantly larger ion $Sr^{2+}$ may also alter the different exchange interactions by making concurrent changes in the Co/Mn-O-Co/Mn bond angles, thereby enhancing the GP.

It is pertinent here to reiterate that the dc magnetization curves showed a sudden slop change forming a smeared peak at lower temperatures which might indicate the presence of concomitant secondary magnetic phases at lower temperature. On the other hand, ac susceptibility measurement is a fine gauge for monitoring the spin dynamics of a system [21]. Again, a comprehensive ac susceptibility study of the both pure and 25% Sr doped PCM systems is hitherto unreported. Hence we have performed the ac magnetization measurements on the present systems to unravel its lower temperature spin dynamics. The Fig. 4(a) and 4(b) are demonstrating the temperature variation of real ($\chi'$) and imaginary ($\chi''$) parts of ac susceptibility data for PCM system. Two clear peaks suggesting two magnetic transitions can be observed at $T_{c1}$~165 K and $T_{c2}$~140 K in both the $\chi'$ and $\chi''$ curves. These two transitions at the same temperatures have been concomitantly observed in the temperature dependent dc (FC) magnetization study as well which is agreeing well with earlier report (Fig. 4c) [39]. The first transition $T_{c1}$ is associated to the FM transition originated due to the super-exchange interactions $Co^{2+}$-$O^{2-}$-$Mn^{4+}$ from the ordered sub-lattice whereas the second transition $T_{c2}$ can be attributed to the FM transition arising due to the super-exchange interactions $Co^{3+}$-$O^{2-}$-$Mn^{4+}$ introduced from the intrinsic disorder of the system [49]. Another smeared peak below ~10 K can also be observed (in $\chi'$) which is seemingly associated to the onset of anti-parallel alignment of paramagnetic $Pr^{3+}$ spins by the internal field of FM component of Co/Mn sub-lattice [64]. This eventually causes sizeable compensation in the FM moment and thus the magnetization drops forming the observed smeared peak. Apart from this, a closer inspection in the $\chi''$ curves at lower temperatures ~38 K interestingly revealed another transition showing frequency dependent broad peaks. The peaks are observed to shift towards the higher temperature with increase in frequency which is typically observed in oxide systems showing short range magnetic interactions. Hence, these dispersive peaks are clearly suggesting towards the presence of slow spin relaxation leading to a re-entrant glassy state near this temperature region. To get further insights into the observed glassy state, we have estimated the Mydosh parameter (p) from the frequency dispersion of the peaks which is a universal tool to distinguish disorder driven SG or CG states from the super-paramagnetic states. The parameter p is defined as the relative change in the freezing temperature $T_f$ per decade shift in the frequency (f) i.e.

$$\boldsymbol{p} = \frac{\Delta \boldsymbol{T_f}}{\boldsymbol{T_f} \Delta \boldsymbol{log_{10}(f)}} \text{ [65]'}$$

For, typical SG or CG systems, p lies between 0.005 and 0.08, while for super-paramagnetic system, it is greater than 0.2. The obtained value of p ~ 0.03 for PCM thus confirms the SG like state at lower temperatures.

Furthermore, the dynamics of the spins in SG or CG states slow down below the critical temperature $T_f$, and it usually follows the dynamic scaling law of the form [27]:

$$\tau = \tau_0 \left(\frac{T_f - T_G}{T_G}\right)^{-z\nu} ;$$

Where $f_0$ denotes excitation frequency corresponding to the characteristic spin-flip time ($\tau_0$) as $f_0=\frac{1}{\tau_0}$ ; $T_G$ represents the equivalent SG freezing temperature for the limit of $H_{DC} \rightarrow 0\ Oe$ and $f \rightarrow 0\ Hz$, zν is called as the dynamical critical exponent. Fig. 4(d) depicts the plot of "$\log_{10}(\tau)$ as a function of $\log_{10}[(T_f\text{-}T_G)/T_G)]$" in its best fit condition which further confirms the re-entrant glassy state in PCM. The best fitting yielded $T_G$~30.6 K, $\tau_0$~$10^{-11}$ s and zν~11.46. For a conventional SG state, $\tau_0$ and zν are typically found to be of ~$10^{-12}$-$10^{-13}$ s and 4-12 respectively. Hence, the above analysis suggests that the system enters in a conventional re-entrant SG state near ~38 K. However, the value of $\tau_0$ is found to be one order bigger than the values typically observed in conventional SG state but again it is quite smaller than the values ~$10^{-6}$-$10^{-9}$ s usually observed in CG states [23]. Eventually, the slightly larger value of $\tau_0$ may indicate towards the existence of some FM spin clusters in the global SG state [66].

To study how the spin dynamics of PCM gets affected by hole ($Sr^{2+}$) doping, we have further performed the ac susceptibility study on PCMS25 (Fig. 5a-b). On looking at the figure, a few interesting features in the ac $\chi'$ and $\chi''$ curves can be readily noted: (1) a mere absence of the first FM transition peak (i.e. $T_{c1}$) is observed. (2) on the contrary, the second FM transition peak associated to the disordered sub-lattice (i.e. $T_{c2}$) is observed to be present at ~153 K. (3) most interestingly, the $\chi''$ curves revealed emergence of a new broad and frequency dispersive glassy peak at ~105 K which was absent in pure PCM system. (4) however, the previously observed re-entrant SG peaks are also observed in PCMS25 system also at ~35 K. (5) the observed peak below ~10 K is again arising due to the onset of $Pr^{3+}$ spin alignment anti-parallel to FM component of Co/Mn sub-lattice [64]. As a matter of fact, reports deciphering the existence of the double-glass magnetic phases in a system are extremely scarce and thus the present observation is very interesting. Eventually, for the disordered Heisenberg systems, theoretical predictions are available for finding two glassy transitions below the long range transition temperature [67,68]. To date, there are only a very few reports available on such systems exhibiting double-glass transitions [69-71].

The dramatic disappearance of first FM peak ($T_{c1}$) can be ascribed to the significant decrease in the $Co^{2+}$ ions as it is evident from the XPS data analysis. Concurrently, it leads to the drastic reduction in the FM $Co^{2+}$-$O^{2-}$-$Mn^{4+}$ interactions which account for the observed disappearance of the $T_{c1}$ peak. In contrast, the sharp and frequency independent peaks at ~153 K can be presumably attributed to the long range FM ordering owing to the $Co^{3+}$-$O^{2-}$-$Mn^{4+}$ interactions as already discussed for PCM. However, to further ascertain the above conclusion, we have recorded the ac susceptibility data of a system (PCMS15) with intermediate (15%) $Sr^{2+}$ doping (Fig. 6 c). It is interesting to note that for this system as well the first FM peak i.e. $T_{c1}$ at ~165 K is observed to be largely suppressed (a closer view is shown in the right inset of Fig. 5c). Thus, it further confirms that $Sr^{2+}$ substitution causes systematic decrease in the FM $Co^{2+}$-$O^{2-}$-$Mn^{4+}$ interactions (due to largely reduced $Co^{2+}$ concentration) and thus in turn leads to the suppression of the $T_{c1}$ peak. Moreover, for this case (PCMS15) the second FM transition $T_{c2}$~140 K is observed to be the most intense peak, suggesting that the long range FM ordering owing to the $Co^{3+}$-$O^{2-}$-$Mn^{4+}$ interactions has been enhanced appreciably. A closer look at the lower temperature (<140 K) features of the $\chi''$ curves reveals that a new broad feature in between the long range FM ($T_{c2}$~140 K) peak and the re-entrant SG peak (~36 K for PCMS15) is arising which was absent in pure PCM system (inset left of Fig. 5c). We can reiterate that a broad feature was also observed for the PCMS25 system. Factually, this intermediate feature was observed at higher temperature for higher doping. However, irrespective of the exact origin of this newly emerged broad peak immediate below $T_{c2}$, it is plausible to state that it is purely associated to the $Sr^{2+}$ substitution. It is pertinent to note here that for both the pure (PCM) and intermediate $Sr^{2+}$ substituted system (PCMS15), the relative intensity of the long range FM transition ($T_{c2}$) was observed to be very high as compared to the lower temperature glassy peaks. However, for higher doping system (PCMS25), these lower temperature features have become more conspicuous as the relative intensity of the $T_{c2}$ got diminished. This is possibly due to the Sr substitution induced effective increase in the different competing AFM interactions as already mentioned above. These different AFM exchange paths progressively weaken the long range FM orderings viz., $T_{c1}$ as well as $T_{c2}$ and thus leading to the gradual suppression of the FM peaks with increased Sr doping. As a consequence, it gradually unveils the lower temperature glassy peaks.

Further investigations with different models have been done on this maidenly recognized Sr doping induced peak (at ~105 K) to get more insights into its nature and origin (Fig. 5b). The estimated value of the Mydosh parameter (p) is found to be ~0.04 (calculated similarly as was done for PCM) which is in the realm of the typical SG or CG state of the oxide systems. This essentially indicates towards the rise of a new SG or CG like state at ~105 K in addition to the glassy state observed at lower temperature ~34 K. Moreover, the critical slowing down of the spin relaxation below $T_f$ has been investigated by dynamic scaling theory as described earlier. Fig. 5(d) depicts the best fit representation of the experimental data (τ dependence of $T_f$) by dynamic scaling model which was achieved with $T_G$~95.2 K, zυ~5.46 and $\tau_0$~$10^{-7}$ s. However, the value of $\tau_0$~$10^{-7}$ s is smaller by few orders than that of the earlier observed SG state in PCM ($\tau_0$~$10^{-11}$ s). The observed value of $\tau_0$~$10^{-7}$ s falls in the range of ~$10^{-6}$-$10^{-9}$ s which is usually found in CG states of oxide systems [23]. In the present case, the larger flipping time is seemingly associated to the slower relaxation of the interacting "spin clusters" instead of individual atomic spins. Again, the smaller value of zυ~5.46 as compared to that of the re-entrant SG state of PCM (~11.46) may also be a possible indication of its CG nature [49]. Therefore, the analysis yields that the Sr doping drives the system into a CG state on lowering the temperature by breaking the long range FM order and aiding the cluster formation.

Again, an attempt to fit the data with Arrhenius law turned out to be unsatisfactory fitting with the unphysical fitting parameters. Thus, it suggests that the underlying phenomena is not simply associated to the thermally activated "single-flip" freezing instead it is essentially a co-operative "spin-cluster" freezing process. Hence, to lend more strength to our analysis, we have used the phenomenological Vogel-Fulcher (V-F) model to fit the same data $T_f$ (τ) as it is generally used for investigating the magnetically interacting spin-clusters [25]. This model has the following form:

$$\tau = \tau_0 \exp\left(\frac{E_A}{K_B(T_f - T_0)}\right);$$

Here, $\tau_0$ represents the characteristic spin-flip time, $T_0$ is the VF temperature which is often termed as the inter-cluster interaction strength and $E_A$ is the activation energy. However, an appropriate approach in fitting the data is crucial otherwise it will result in wrong parameters. Thus, we have followed the method of Souletie and Tholence in estimating the proper value of $T_0$~90.7 K [72]. This value is then used to make the linear fit of the plot "ln(τ) Vs $1/T_f - T_0$" to determine the other parameters (Fig. 5(e)). The fit yielded the values of $\tau_0$ to be~$10^{-7}$ s and the activation energy $E_A/K_B$ to

be~113.5 K which are quite reasonable. Eventually a non-zero value of $T_0$ suggests towards the presence of inter-cluster interactions [72]. In the frame of V-F model, $T_0$<< $E_A/K_B$ signifies weak inter-cluster interactions whereas $T_0$>> $E_A/K_B$ indicates that to be strong. For the present case, $T_0$~0.8 $E_A/K_B$ suggests for the intermediate inter-cluster interactions. Again the larger value of $\tau_0$ also supports the rise of interacting spin-clusters. Therefore, all the above analysis equivocally confirms that the observed transition is in the realm of the CG state.

Further, to investigate whether the nature of the glassy peak observed at lower temperature ~34 K for PCMS25 is similar to that of the earlier observed SG peak at ~38 K in pure PCM system, we have analyzed this peak by different model fittings. The frequency sensitivity has been measured by estimating the Mydosh parameter (p) which is found to be ~0.033. This value of p corroborates well with that of the typical SG or CG states. Fig. 5(f) demonstrates the best fit line with dynamic scaling law for the "τ dependence of $T_f$" data of 34 K peak. The best fit yielded the following parameters: $T_G$~27 K, zυ~12.25 and $\tau_0$~$10^{-11}$ s. In contrast to the higher temperature CG peak at ~105 K, the analysis of this lower temperature (~34 K) peak yielded a higher value of zυ and smaller value of $\tau_0$ by few orders which are qualitatively similar to that of the re-entrant SG state of pure PCM system. Therefore, the lower temperature peak can be associated to the typical freezing of the individual atomic spins instead of the spin-clusters. Again, it is relevant to note that the observed $\tau_0$ value is higher by an order than that of typical SG states (~$10^{-12}$-$10^{-13}$ s). This is seemingly associated to the effect of the existence of some residual FM clusters due to the higher temperature CG state (~105 K). However, observation of the similar re-entrant SG states in the same temperature range for both the pure (PCM) and doped (PCMS25) systems indicates that they are of same origin which is the inherent ASD driven spin frustration as already discussed.

One of the important characteristics of an ordinary SG or CG system is the applied dc field dependent downward shifts of the freezing temperature ~$T_f$ which eventually get smeared out with sufficiently high field due to domain wall relaxation [73]. Fig. 6(a) depicts the $\chi''$(T) data for PCMS25 system at different dc magnetic fields. It is evident from the figure that the $\chi''$ curves get shifted towards lower temperatures with increasing magnetic fields for both the peaks (SG~34 K and CG~105 K) which is consistent with the glassy nature (unlike the FM $T_{c2}$ peak which shifts towards higher temperature) [66,73]. In the conventional SG or CG systems, the $T_f$(H) in low fields usually follows the de Almeida-Thouless (A-T) line which is given by the following equation [74]:

$$H_{dc}^{2/3}=\Delta J^{2/3}\left[1-\frac{T_f(H)}{T_f(0)}\right]$$

Here, $T_f$(0) represents the freezing temperature in the absence of any applied dc magnetic fields and ΔJ is a constant. It is evident from Fig. 6(b) that our experimental data of the CG peak has shown satisfactory fit with the above equation. It yielded the value of $T_f$(0) to be ~104.2 K which is an excellent match with our experimental data. Hence, it provided further confirmation to identify the intermediate peak (~105 K) as an onset of CG state. However, the $T_f$ of the lower temperature SG peak showed very subtle shift with the fields and thus could not satisfactorily meet the above A-T line criterion. This feeble field dependence of $T_f$ in similar re-entrant SG systems is however common [75].

Moreover, further confirmation of the glassy state can be achieved by examining the isothermal temporal relaxation of the residual magnetization m(t) which is an experimental realization of the nonergodicity present in the spin-frustrated system [24]. Hence, in performing this experiment, we have first cooled the PCMS25 system with an applied field of 100 Oe from 300 K down to different desired temperatures in the CG and SG regimes. After waiting for 100 s, the residual magnetization data have been recorded for a sufficiently long period of time after switching off the field. Fig. 6(c) is showing the time evolution of the normalized residual magnetization m(t)=$\left(\frac{M_t}{M_{t=0}}\right)$ at three different temperatures 90 K, 80 K and 70 K in the CG regime. It is evident from the above figure that even after such a long time of 4500 s, the decay in the residual magnetization continues which is a typical aspect of disordered glassy states owing to the existence of the hierarchical arrangements of the meta-stable states [24,25]. Typically, such isothermal relaxation data of disordered systems is analyzed by using the phenomenological model called KWW (Kohlrausch Williams Watt) stretched exponential equation of the form [24]:

$$m(t) = m_0 - m_g exp\left\{-\left(\frac{t}{\tau}\right)^{\beta}\right\};$$

Here, $m_0$ and $m_g$ are representing the initial residual magnetization and glassy component's magnetization respectively, $\tau$ represents the time constant related to the characteristic relaxation and $\beta$ is termed as shape parameter which lies in between 0 and 1 in conventional SG states. However, an attempt to fit all the three data with the KWW model resulted in unsatisfactory fittings. In fact, the data deviated from KWW model in the sharply decaying m(t<250 s) region while it is observed to show reasonably good fit in the higher time region (t>250 s) (only the KWW fit of 90 K data is shown in inset of Fig. 6c). An alternative attempt to fit the data was made using the logarithmic relaxation usually observed in the long range spin ordered regimes which is of the form: $m(t)=m_0-Cln(t)$ [76]. Here, $m_0$ denotes the initial remanent magnetization while C is a material dependent constant. However, the fit using the above model also yielded very poor results (not shown).

The initial rapid down fall of the m(t) data can be presumably attributed to the existence of interacting FM clusters [28]. Eventually, for interacting magnetic particles, Ulrich et. al. have theoretically demonstrated that the decay rate of the spin relaxation:$W(t) = \frac{d[\ln m(t)]}{dt}$ follows a power law given by: $W(t)=At^n$, where A is a temperature dependent prefactor and n is the density dependent critical exponent which also depend on the strength of the magnetic interactions [77]. In this scenario, the remanent magnetization can be expressed by the following equation used by Ulrich et. al [77,78]:

$$m(t) = \frac{M(t)}{M(t_0)} = e^{-\alpha_n}\left[1 + \alpha_n\left(\frac{t}{t_0}\right)^{1-n}\right];$$

Here, $\alpha_n = A\frac{t_0^{1-n}}{|1-n|}$ and m($t_0$) represents initial magnetization at time $t_0$. Finally, the attempts to fit the data with the above equation yielded satisfactory results for all the three temperatures in CG regime as can be seen in Fig. 6(c). Further, the value of the exponent n is observed to increase progressively from 1.363 to 1.521 with increase in temperatures (70 K to 90 K) which is essentially pointing out the presence of the CG state [79]. The observed increase in n is a manifestation of the increasing inter-cluster interactions with increasing temperature [79]. On the contrary, as evident from the Fig. 6(d), the lower temperature ~25 K relaxation data (recorded following the same protocol as done in the CG region) in the re-entrant SG state showed quite reasonable fit with the KWW model. The fit yielded β ~0.449 which is corroborating well with the typical values shown by the conventional disordered SG systems(0<β<1) [24,66]. Hence, all the above analysis eventually suggests that the Sr doping in PCM effectively introduces a new CG like state in the system. Interestingly, the lower temperature re-entrant SG like state arising from the inherent ASD present in the system is found to be quite robust as it did not smear out even after 25% of Sr substitution. Only noticeable difference is that the SG freezing temperature $T_f$ showed a mild shift from 38 K to 34 K.

To elucidate the origin of the observed re-entrant glassiness in PCM, we must consider the role of the anti-site disorder (ASD) which is a potential source for emanating the SG state by hindering the long range ordering [8,11,28,80]. The presence of ASD gives rise to the random exchange bonds formation leading to the spin frustration. The domain formation in a purely FM/AFM system involves microscopic time scales whereas the presence of ASD causes the pinning of the domain wall which essentially produces the metastable states [80]. This eventually allows the domain walls in reaching from one state to the other via thermally activated hopping and thus hindering the system to attain an equilibrium state in the experimental time scale [80]. In addition, unlike the perfectly ordered DP containing only $Co^{2+}$/$Mn^{4+}$ ions situated in two distinct sites, PCM is found to contain additional $Co^{3+}$/$Mn^{3+}$ ions. Eventually, the existence of different competing exchange interactions viz., the existing predominant FM $Co^{2+}$-$O^{2-}$-$Mn^{4+}$ interactions along with the other FM $Co^{3+}$-$O^{2-}$-$Mn^{4+}$ and AFM $Co^{3+}$-$O^{2-}$-$Co^{3+}$, $Co^{3+}$-$O^{2-}$-$Co^{2+}$, $Co^{2+}$-$O^{2-}$-$Co^{2+}$, $Mn^{3+}$-$O^{2-}$-$Mn^{3+}$, $Mn^{4+}$-$O^{2-}$-$Mn^{4+}$ interactions is leading to the inhomogeneous magnetic order through these multiple exchange paths. In this scenario, on lowering the temperature, increasing competition between the aforementioned FM/AFM interactions drive the system into a highly spin frustrated regime weakening the average FM order and finally it re-enters in a SG state owing to the partial random, non-collinear freezing of the spins. Consequently, the system re-enters in a SG like state. Apart from this lower temperature SG like state, the additional higher temperature CG like state is solely associated to the enhanced competition between the different increasing AFM interactions with the existing FM interactions triggered by Sr substitution. Eventually, Sr substitution appreciably increased the $Co^{3+}$ concentration at the cost of the decrease in $Co^{2+}$ ions which significantly enhanced AFM $Co^{3+}$-$O^{2-}$-$Co^{3+}$, $Co^{3+}$-$O^{2-}$-$Co^{2+}$ interactions leading to the breaking of the long range FM ordering into FM clusters. This essentially raised the CG like state below the FM transition $T_{c2}$.

Eventually, the co-existence of multiple inhomogeneous magnetic phases in a system viz., FM/Spin glass, FM/Ferrimagnet, FM/AFM, hard/soft phases of FM systems etc often lead to the exhibition of the exchange bias (EB) [34-37]. Interestingly, the system PCMS25 holds different magnetic phases such as FM/AFM, CG and SG states which provoked us to perform the EB experiment of this sample. We have followed two protocols for estimating the EB effect: (1) spontaneous EB (SEB) has been measured by recording the M(H) loop after cooling the sample from 300 K to 5 K in the absence of dc bias field. (2) conventional EB (CEB) has been measured similarly but in this case a dc bias field of ~+/-5 T has been applied while cooling. Fig. 7(a) is showing the SEB and CEB loops at 5 K. The clear evidences of giant EB effect for both the protocols can be observed as a large horizontal shift of the loops (inset of Fig. 7a). The measurement of the loop asymmetry gives the quantitative value of the EB as $H_E= -(H_L+H_R)/2$ while the coercivity is estimated as $H_C= -(H_L-H_R)/2$, here $H_L$ and $H_R$ are denoting the coercive fields of the left and right of the loop respectively. The observed value of $H_E^{SEB}$ and $H_C^{SEB}$ for SEB measurement are found to be ~0.2053 T and 0.236 T which are quite high. It is relevant to note here that pure PCM system was reported to exhibit CEB of $H_E$~0.5 T but no sizeable SEB was observed for this system [39]. Thus, the observation of such a high SEB for the PCMS25 system can be of particular interest. The observation of the M(H) loop shift towards negative (positive) field (H) direction on cooling the sample with positive (negative) field clearly rules out the possibility of the minor loop effect and thus establishing the CEB effect in the system (Fig. 7a) [81]. For the positive (+5 T) magnetic field (p-type), the CEB effect showed the values of $H_{EP}$ and $H_C^P$ to be ~0.9039 T and 0.1996 T respectively. Again, for the negative magnetic field (-5T) cooling (n-type), the observed CEB effect yielded the values $H_{EN}$ and $H_C^N$ to be~0.8904 T and 0.1820 T respectively. Eventually, the observation of such high values of $H_E$ is comparable with other giant EB systems [82]. However, the observation of the asymmetry in the CEB fields i.e. $|H_{EP}| > |H_{EN}|$ may be arising due to the inherent anisotropy of the system. Further, we have recorded the M(H) loops at two other different temperatures 10 K and 50 K with zero field cooling to probe the impact of temperatures on the SEB effect (Fig. 7c and 7d). The $H_E^{SEB}$ at 10 K has showed a drastically diminished value of ~0.028 T while no sizeable SEB effect has been observed at 50 K. This is seemingly associated to the enhanced AFM ordering below ~10 K which was evidenced by both dc and ac magnetization curves. Moreover, the asymmetry in the remanence of the M(H) curves can also be clearly observed. Interestingly, even after reaching a very high negative magnetic field (-7 T), the remanence remained positive (at H=0 Oe) for the SEB and p-type CEB case. Similarly, the opposite scenario can be observed for the n-type CEB case. This is also a characteristic feature of EB systems [81].

One of the important aspects of EB systems is the training effect (TE) which is realized as a monotonous decrease in the EB field ($H_E$) on field-cycling the system through successive tracing of hysteresis loops with increasing loop index number (n) [83]. Eventually, investigation of the TE is very crucial for the materials exhibiting EB effect since it can effectively monitor the spin dynamics at the interface of two different magnetic structures. A small value of TE is desirable for its potentially wide applicability. We have carried out the TE experiment by consecutive tracing of n=9 hysteresis loops at 5 K after cooling the system (PCMS25) under magnetic field of ~+7 T. In fact, the $H_E$ relaxation showed more prominent variation against field cycling for the left branch of the M(H) loops than that for the right branch which is however common in SG/FM systems. It can be elucidated by the stronger thermally activated (over the anisotropy barrier) domain reversals at the left branch (due to higher magnitude of the sweep field) than that of the right branch [84]. Fig. 7(b) demonstrates the measured TE in CEB mode (for the left branch) where the remanence asymmetry relaxation is clearly visible with increasing field cycling. However, a decrease of only~4% in $H_E$ is observed while tracing the M(H) loops from n=1 to 9. This small decrease in $H_E$ is insignificantly small as compared to the giant value observed in CEB effect. This is a clear indication towards the reasonably stable interfacial pinned spin configuration in the present system [37]. It is often observed that the mutual relationship between the $H_E$ and n can be expressed by the following empirical power law which was coined for the materials exhibiting EB effect [81]:

$$(H_E - H_\infty) \propto \frac{1}{\sqrt{n}};$$

Here, the term $H_\infty$ represents the value of $H_E$ at n→∞. However, the above equation holds only for n>1. Fig. 7(e) demonstrates the reasonably good fit (solid line) of "$H_E$ variation with increasing n" using the above power law. The fit yielded quite a high value of $H_\infty$=0.8636 T for the present system.

The observed TE can be realized on the basis of the relaxation or demagnetization of the pinned moments (formed due to the exchange anisotropy) at the interface of the existing FM/AFM regions by the switching of the field induced soft FM domains back and forth under the influence of the applied loop tracing fields (H).

The observed EB effect in AFM/FM/SG phase separated systems can be explained on the basis of the field induced FM cluster formation in the AFM matrix [81]. Hence, to further ascertain the growth of the FM clusters, we have recorded the temporal evolution of the magnetization at 5 K. The sample was cooled down to 5 K in absence of applied magnetic field. The time variation of the magnetization m(t)=M(t)/M(0) was recorded with two different applied fields 4 T and 7 T after waiting for 100 s (Fig. 7f). It is evident from the figure that for the smaller applied field of ~4 T, the m(t) data shows clear relaxation even after a long time of ~6000 s which is typical feature of a glassy system. However, when the applied field was sufficiently high ~7 T, the m(t) relaxation has been significantly lowered. This is in turn suggesting towards the progressive growth of the FM clusters at the cost of the glassy spins with increasing fields [85].

To elucidate the observed exchange bias effect in the present material, a phenomenological model comprising of field induced soft FM small clusters embedded in the global disordered AFM matrix can be discussed [81]. As already discussed, the present system PCMS25 is a phase separated system consisting of multiple magnetic phases: FM/AFM/glassy states. The co-existence of these multiple magnetic phases is also evident from the M(H) curves which did not saturate even after reaching a field of 7 T. The magnetization yielded a maximum value of ~1.7 $\mu_B$/f.u at +7 T which is much smaller than the saturation magnetization expected for the perfectly Co/Mn ordered ferromagnet (6 $\mu_B$/f.u) [39]. The hole doping in the present system raised the dominant AFM interactions breaking the long range FM ordering and ultimately lead to the glassy states at lower temperatures. In fact, for the present system, even it enters in a glassy phase (CG/SG), the FM and AFM interactions co-exist (with predominant proportion being AFM) with the partially frozen glassy state which is also a characteristic of re-entrant SG/CG systems [26]. To interpret the CEB effect schematically, we have considered a model of an AFM core which is surrounded by frozen glassy shell as shown in Fig. 7(g) [36]. During the field cooling, the glassy spins align along the strong applied field thus forming a super-ferromagnetic (SFM) region inside the AFM matrix. The asymmetry of M(H) curves thus can be elucidated by the rise of an interface magnetic moment ($m_i$) (due to the exchange interactions between the field induced SFM clusters and existing AFM phase) which remains frozen due to the kinetic arrest at low temperatures [81]. This frozen surface moment $m_i$ eventually provides the required exchange anisotropy or the pinning forces to the FM domains which give rise to the positive (negative) remanence during the recoiling M(H) curve for the p-type (n-type) CEB case [81]. This in turn produces a horizontal shift of M(H) curves leading to the observed EB effect.

## Conclusion

Summarizing, we have carried out a thorough investigation of the hole substitution ($Sr^{2+}$) induced alterations in the electronic structures, transport, dc and ac magnetization properties of a partially B-site disordered double perovskite $Pr_{2-x}Sr_xCoMnO_6$. The room temperature (300 K) XPS data analysis suggested for the existence of a mixed valence nominal oxidation states of the B-site ions ($Co^{2+}$/$Co^{3+}$ and $Mn^{3+}$/$Mn^{4+}$). The XPS data analysis further showed an effective enhancement in the average valence states by the hole doping. The XPS valence band spectra analysis showed the mere absence of the electronic states near the Fermi level ($E_F$) for both the pure and doped systems which in turn suggest for the insulating nature of them. However, $Sr^{2+}$ substitution is observed to cause little enhancement in the spectral weight near $E_F$, thus predicting for an enhanced conductivity. The electrical resistivity study revealed an insulating ground state for both the systems and thus corroborated the VB spectra results. Both the resistivity data satisfactorily followed the VRH mechanism in the entire temperature range. Moreover, the analysis showed that the hole doping essentially increased the carrier concentration which was also reflected from the VB spectra study as already mentioned. At lower fields, the temperature variation of the inverse susceptibility curves of the hole doped sample PCMS25 divulged the signature of a typical Griffiths like phase which was further confirmed by a power law analysis. The plausible origins from which the observed GP can emanate include: the inherent B-site disorder along with the disorder related to the random substitution of Sr at Pr-site, the concurrent alterations in the Mn-O-Mn bond angles due to smaller Sr substitution and quenched disorder arising from the J-T active ions. Moreover, the ac susceptibility studies for both the pure and doped systems maidenly recognized a RSG state below ~40 K which can be presumably attributed to the spin frustration arising from the inherent ASD along with the presence of different charge states of the B-site ions. Most interestingly, apart from this lower temperature SG state (<40 K), an additional glassy state immediately below the long range FM ordering ($T_{c2}$) has been observed to appear exclusively due to the hole doping. Furthermore, the analysis of the ac susceptibility as well as the temporal spin relaxation data by different model fits equivocally established existence of the two distinct glassy states: the lower temperature SG state and the higher temperature CG state in the hole doped system. The onset temperature of the CG state is observed to increase with the increased Sr doping. Eventually, the increasing competition between the FM/AFM interactions owing to the different exchange paths (triggered by hole doping) on lowering the temperature breaks the long range FM ordering into clusters leading to the CG state. However, to date, no comprehensive report is available deciphering the co-existence of dual glassy states in double perovskites materials. In fact, observation of such double disordered glassy transitions is extremely scarce which essentially places the present system amongst the rare materials. Moreover, the hole doping lead to the exhibition of giant spontaneous and conventional EB effect. A model of field induced soft FM clusters embedded in the global AFM matrix is invoked to explain the observed EB effect. Thus, the present system (PCMS25) exhibited a spectrum of potentially important magnetic properties which is also very rare to observe.

**Acknowledgements**

The authors are thankful to the Central Instrumentation Facility Centre, Indian Institute of Technology (BHU) for providing the magnetic measurement facilities (MPMS).

**Figure captions:**

**Fig. 1**: (a) XPS survey scan of PCM (top) and PCMS25 (bottom). (b), (c), (e) and (f) are depicting core level Pr*3d*, O*1s*, Mn*2p*, Co*2p* XPS spectra for PCM (top panel) and PCMS25 (bottom panel) respectively. (d) represents Sr*3d* core level XPS of PCMS25. (g) show VB spectra at 300 K for PCM (top) and PCMS25 (bottom).

**Fig. 2**: Resistivity (ρ) vs temperature (T) plots for PCM and PCMS25 systems along with their VRH fits.

**Fig. 3**: M(T) ZFC-FC curves at 100 Oe for PCMS25. Inset top shows "dM/dT vs T" plot. Inset bottom presents "Curie-Weiss fit to the $1/\chi$ vs T" plot.

**Fig. 4**: (a) and (b) show ac $\chi'(T)$ and $\chi''(T)$ curves respectively for PCM. Inset of (b) presents enlarged view of $\chi''(T)$ near SG region. (c) depicts dc FC M(T) curve for PCM. (d) demonstrates the dynamic scaling fit of the $T_f(\tau)$ data.

**Fig. 5**: (a) and (b) show ac $\chi'(T)$ and $\chi''(T)$ curves respectively for PCMS25. (c) depicts the $\chi''(T)$ curves for PCMS15 while its insets are showing the closer views of its shaded regions. (d) and (e) are presenting the dynamic scaling and V-F fits to the "$\tau$(=1/f) dependence of $T_f$" data in the CG region for PCMS25. (f) shows the dynamic scaling fit of $T_f(\tau)$ data in the SG region for PCMS25.

**Fig. 6**: For PCMS25:(a) presents $\chi''(T)$ curves at different fields (H) (b) shows the A-T line fit to the "$T_f$ vs $H^{2/3}$" plot. (c) depicts the m(t) data at 90 K, 80 K and 70 K along with their fits with Ulrich's equation. (d) KWW fit to the m(t) data recorded at 25 K.

**Fig. 7**: For PCMS25: (a) shows the M(H) curves with cooling fields of 0 T and +/-5 T. Its inset depicts the enlarged view. (b) presents the enlarged section of the left branches of the n=9 field cycled M(H) loops for TE. (c) and (d) shows the M(H) loops at 10 K and 50 K respectively. (e) depicts the power law fit to the "$H_E$ vs n" curve. (f) shows "M(t)/M(0) vs t" plot at 5 K with field 4 T while its inset shows that with fields 4 T and 7 T. (g) demonstrates the core-shell model with AFM core and glassy shell for EB effect.

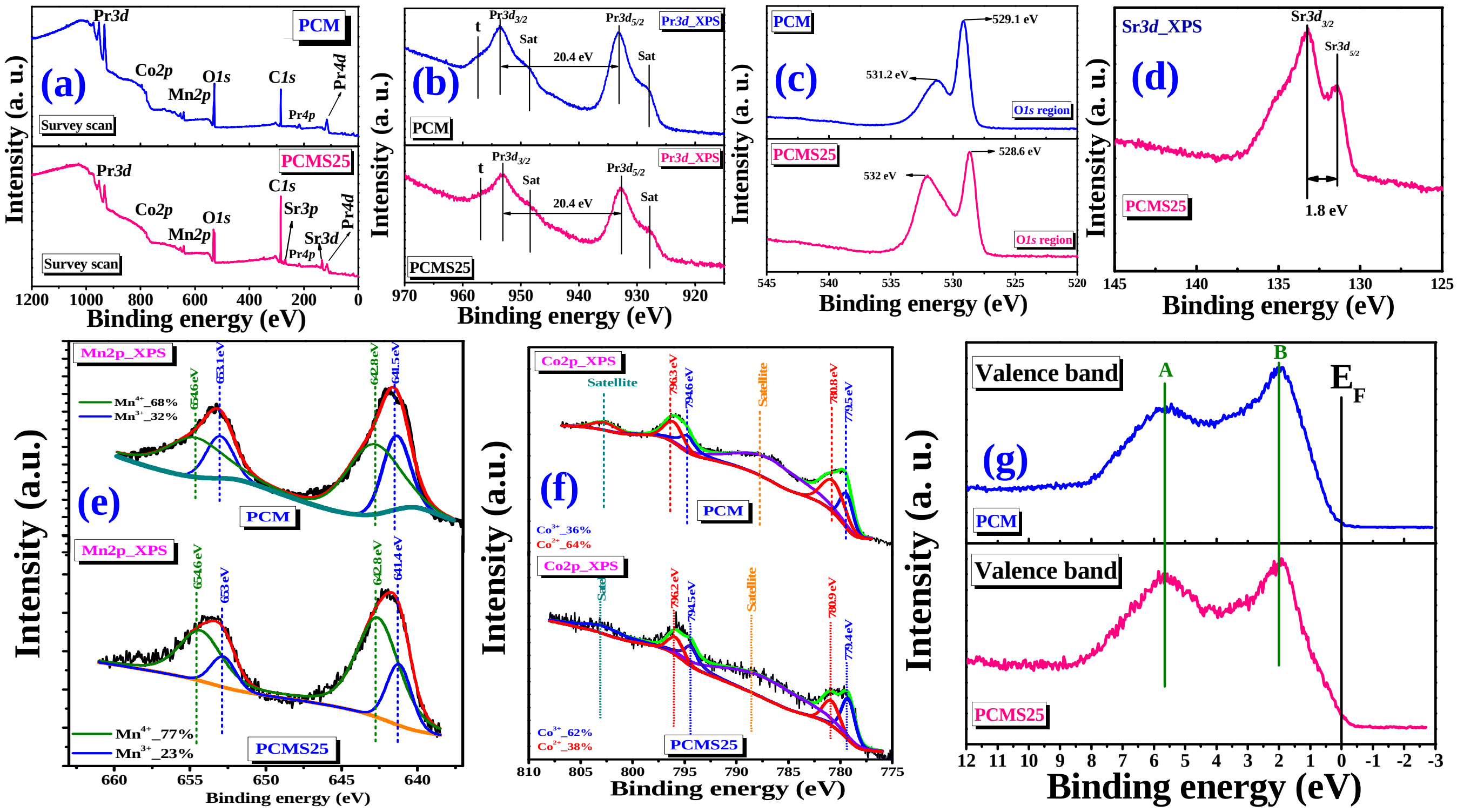

Figure 1

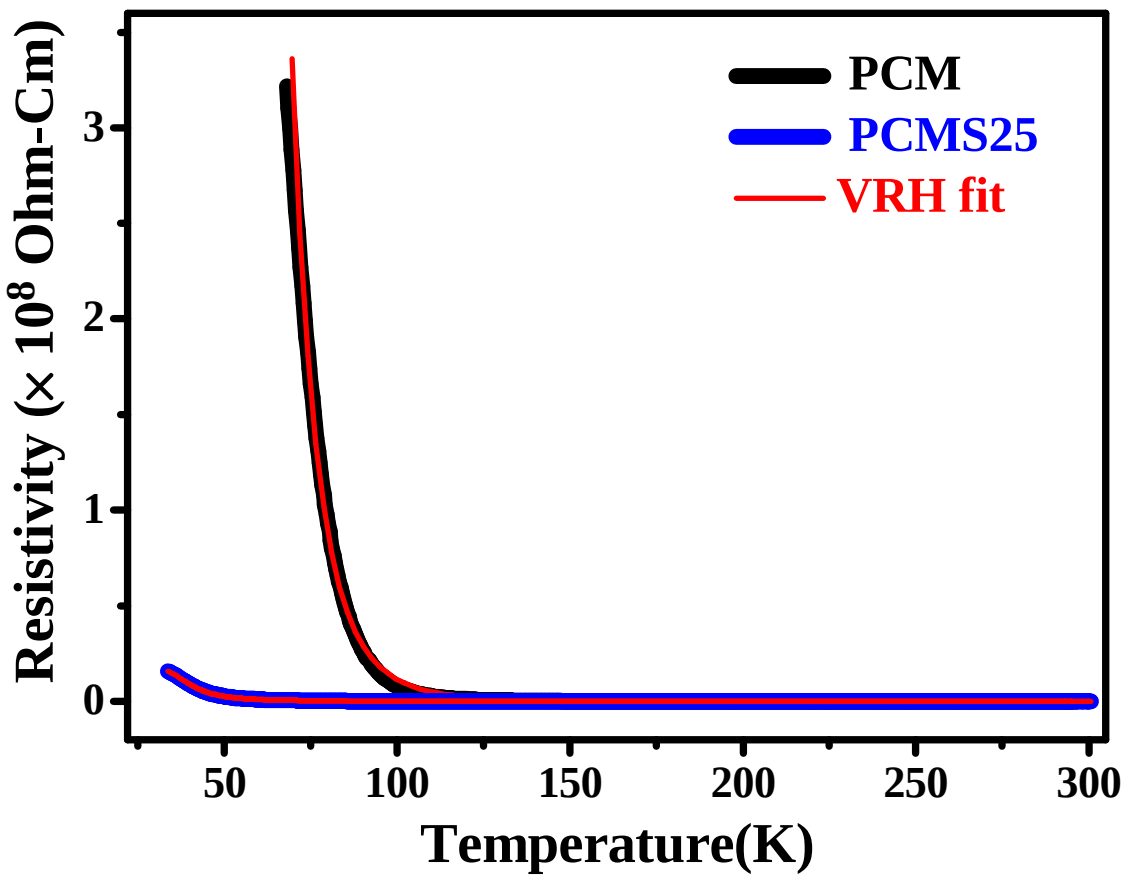

Figure 2

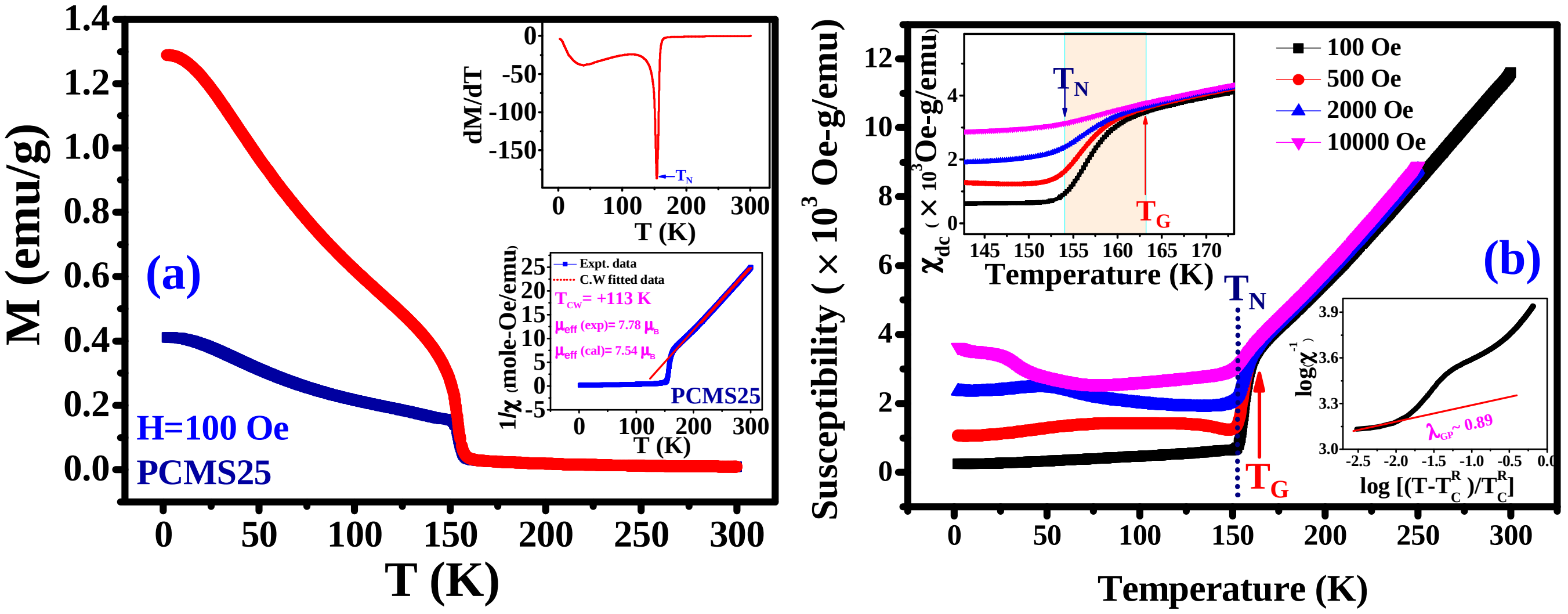

Figure 3

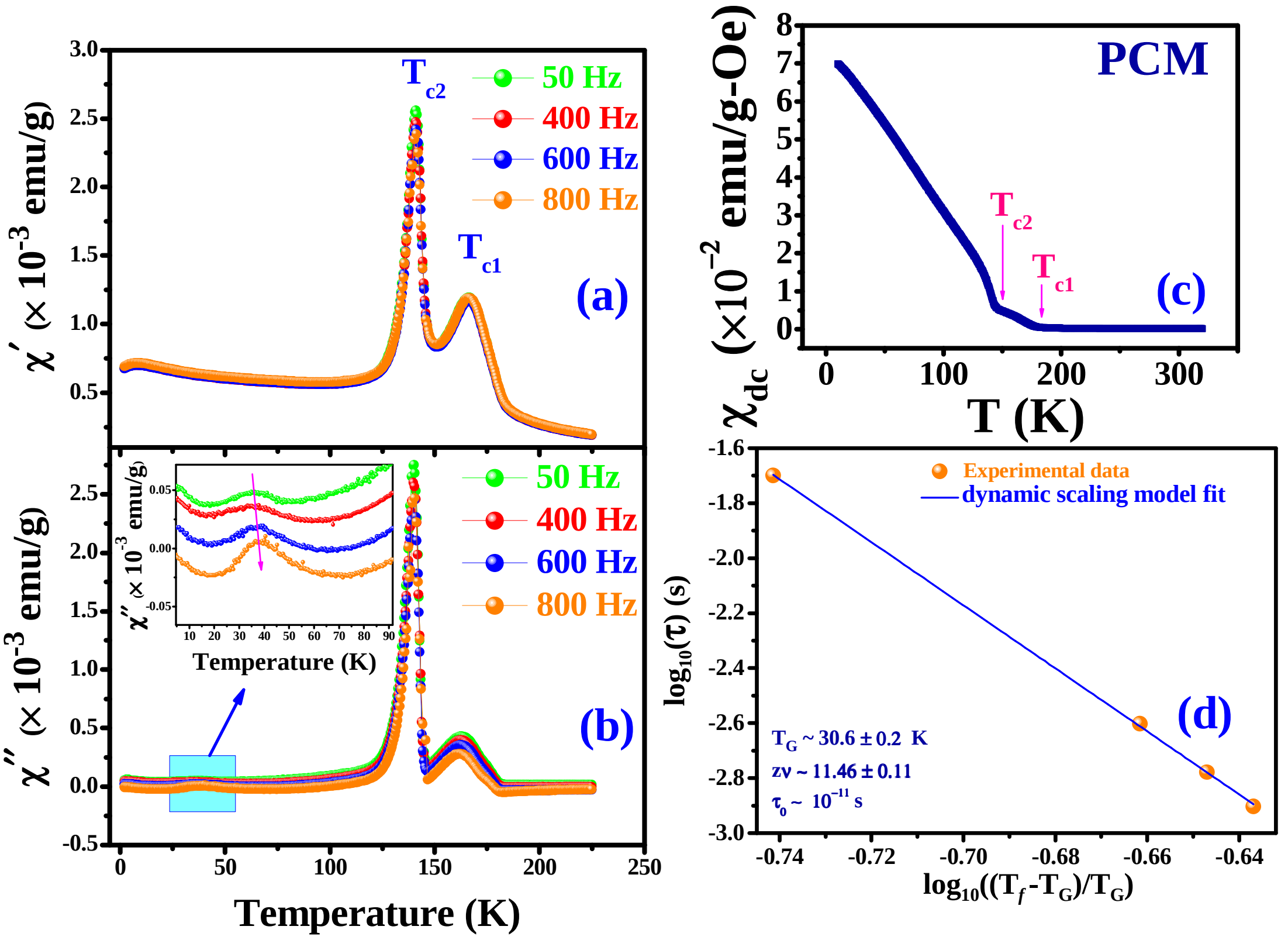

Figure 4

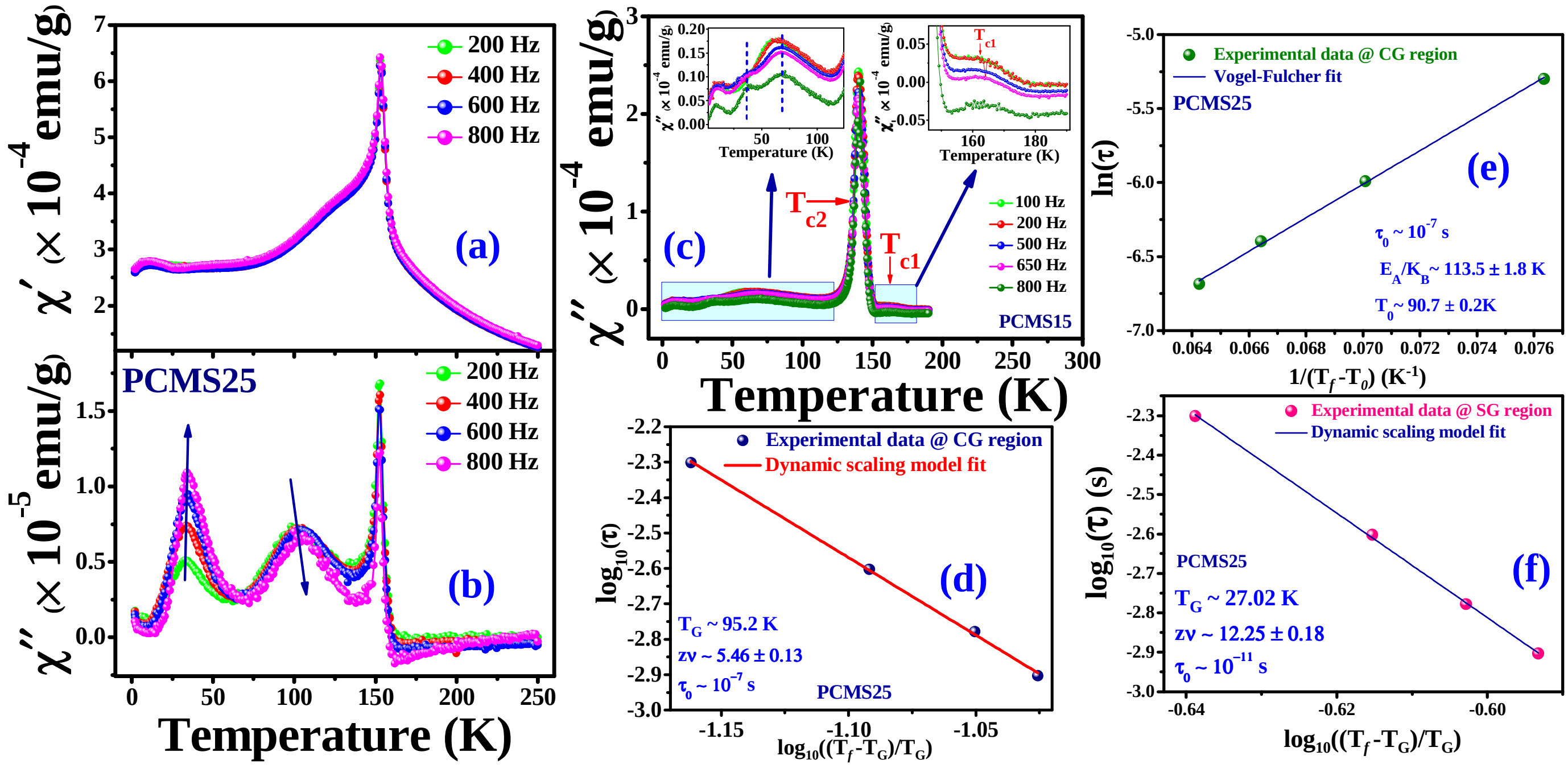

Figure 5

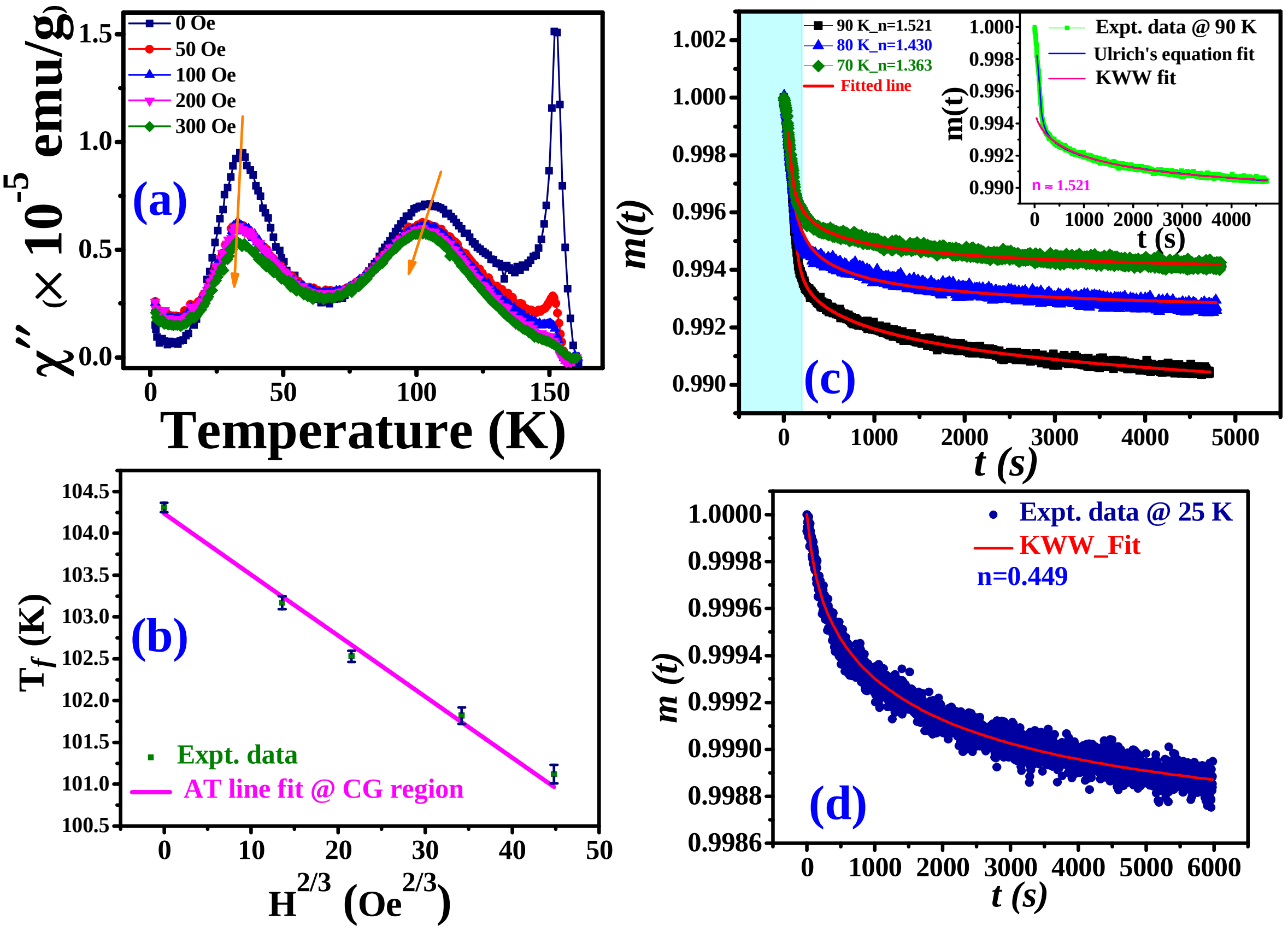

Figure 6

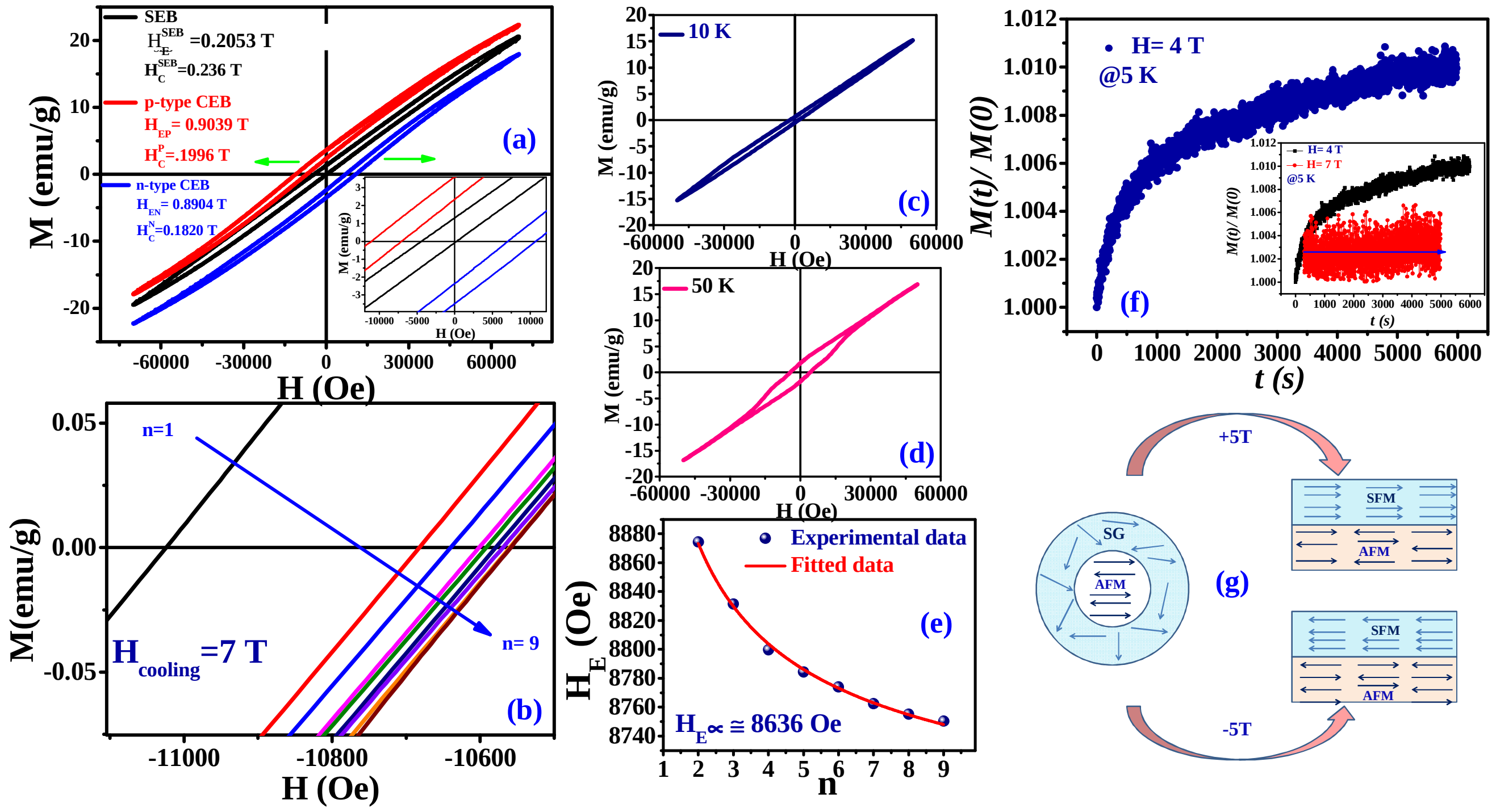

**Figure 7**